\documentclass[nohyper,12pt,letterpaper]{JHEP}
\usepackage[dvips]{epsfig}
\usepackage{amsfonts,amssymb}

\newcommand{\be}{\begin{equation}}
\newcommand{\ee}{\end{equation}}
\newcommand{\bea}{\begin{eqnarray}}
\newcommand{\eea}{\end{eqnarray}}
\newcommand{\ba}{\begin{array}}
\newcommand{\ea}{\end{array}}

\def\bbox{{\,\lower0.9pt\vbox{\hrule \hbox{\vrule height 0.2 cm
\hskip 0.2 cm \vrule height 0.2 cm}\hrule}\,}}
\newcommand{\dsl}{\pa \kern-0.5em /}

\newcommand{\EQ}{\begin{equation}}
\newcommand{\EN}{\end{equation}}

\def\bbox{{\,\lower0.9pt\vbox{\hrule \hbox{\vrule height 0.2 cm
\hskip 0.2 cm \vrule height 0.2 cm}\hrule}\,}}

\newcommand{\pa}{\partial}

\def\be{\begin{equation}}
\def\ee{\end{equation}}
\def\ba{\begin{eqnarray}}
\def\ea{\end{eqnarray}}

\def\bq{\begin{quote}}
\def\eq{\end{quote}}

\makeatletter
\def\vereq#1#2{\lower3pt\vbox{\baselineskip1.5pt \lineskip1.5pt
\ialign{$\m@th#1\hfill##\hfil$\crcr#2\crcr\sim\crcr}}}
\makeatother
 at 10truept

\newcommand{\beq}{\begin{equation}}
\newcommand{\eeq}{\end{equation}}
\newcommand{\beqa}{\begin{eqnarray}}
\newcommand{\eeqa}{\end{eqnarray}}

\def\ltap{\ \raise.3ex\hbox{$<$\kern-.75em\lower1ex\hbox{$\sim$}}\ }
\def\gtap{\ \raise.3ex\hbox{$>$\kern-.75em\hbox{$\sim$}}\ }
\def\gl{\ \raise.5ex\hbox{$>$}\kern-.8em\lower.5ex\hbox{$<$}\ }
\def\roughly#1{\raise.3ex\hbox{$#1$\kern-.75em\lower1ex\hbox{$\sim$}}}

\title{Quantum Black Holes as Holograms in AdS Braneworlds}

\author{Roberto Emparan\thanks{Also at Departamento de F\'\i sica
Te\' orica, Universidad del Pa\'\i s Vasco, Bilbao, Spain.} \\
    Theory Division, CERN, CH-1211 Geneva 23, Switzerland \\
E-mail: \email{{\tt roberto.emparan@cern.ch}}}
\author{Alessandro Fabbri\\
   Dipartimento di Fisica dell'Universit\`a di Bologna $\&$ INFN sezione di Bologna\\
   Via Irnerio 46, 40126 Bologna, Italy \\
E-mail: \email{{\tt fabbria@bo.infn.it}}}
\author{Nemanja Kaloper\\
   Department of Physics, Stanford University, Stanford, CA 94305-4060, USA \\
E-mail: \email{{\tt kaloper@stanford.edu}}}

\abstract{ We propose a new approach for using the AdS/CFT
correspondence to study quantum black hole physics. The black
holes on a brane in an AdS$_{D+1}$ braneworld that solve the
classical bulk equations are interpreted as
duals of {\it quantum-corrected} $D$-dimensional black holes, rather
than classical ones, of a conformal field theory coupled to gravity.
We check this explicitly in $D=3$ and $D=4$.  In $D=3$ we reinterpret
the existing exact solutions on a flat membrane as states of the dual
$2+1$ CFT. We show that states with a sufficiently large mass
really are $2+1$ black holes where the quantum corrections dress the
classical conical singularity with a horizon and censor it from the
outside. On a negatively curved membrane, we reinterpret the classical
bulk solutions as quantum-corrected BTZ black holes. In $D=4$ we argue
that the bulk solution for the brane black hole should include a
radiation component in order to describe a quantum-corrected black hole
in the $3+1$ dual. Hawking radiation of the conformal field is then
dual to classical gravitational bremsstrahlung in the AdS$_5$ bulk.}

\preprint{SU-ITP-02/23\\ CERN-TH/2002-131 \\ \tt{hep-th/0206155} \\
}

\begin{document}

\section{Introduction}

We propose here a connection between two seemingly unrelated
problems in black hole theory: {\it i)} the well-known problem of
the backreaction from quantum effects on a black hole geometry,
and {\it ii)} the description of a black hole in an AdS
braneworld, as in the Randall-Sundrum model with an infinite extra
dimension, RS2 \cite{rs}. Quantum fields in a black hole
background lead to particle production and black hole evaporation
via Hawking radiation \cite{swh}. To leading order in perturbation
theory, this yields an expectation value of the renormalized
stress-energy tensor of quantum fields $\left<T_{\mu\nu} \right>$,
which includes quantum corrections. The backreaction of
$\left<T_{\mu\nu} \right>$ on the classical geometry modifies it
according to the one-loop corrected Einstein's equation
$G_{\mu\nu}= 8\pi G_4\left<T_{\mu\nu}\right>$. Unfortunately, the
stress-energy tensor $\left<T_{\mu\nu} \right>$ in a black hole
spacetime can only be computed approximately, while determining
its backreaction is even more difficult \cite{books}. Only in
dimensions $D<4$ was it possible to find exact solutions
\cite{steif,lo,shm,stro}.

On the other hand, an AdS braneworld consists of a bulk AdS$_{D+1}$
space ending on a $D-1$-dimensional domain wall, or brane. A prototype
is the RS2 model where AdS$_5$ ends on a $3$-brane, which should model
our 3+1 dimensional world. It is therefore natural to look for a
suitable description of a black hole in this scenario. However, the
attempts to find exact, static, asymptotically flat black hole
solutions localized on the brane in AdS$_{D+1>4}$, with regular
horizons both on and off the brane, have come empty-handed to date (for
published examples see, e.g., \cite{chr}-\cite{cfm}). It has even been
suggested that static, asymptotically flat, spherical black holes on
the brane might not altogether exist in the RS2 model
\cite{bgm}\footnote{Ref.~\cite{wiseman} obtains a numerical solution
for a static star on an RS2 brane.}. Contrasting this, exact static
solutions localized on a 2-brane in AdS$_4$ have been found in
\cite{ehm1,ehm2}.

Here we adopt the point of view that the difficulties in
constructing these solutions are no mere accident, but are
intricately related to the effects induced by quantum corrections.
We use a modification of AdS/CFT correspondence \cite{malda} for
the RS2 model \cite{apr1}-\cite{apr7} to connect both problems.
Our main result is the following conjecture:

\begin{quotation}
\noindent {\it The black hole solutions localized on the brane in
the AdS$_{D+1}$ braneworld which are found by solving the
classical bulk equations in AdS$_{D+1}$ with the brane boundary
conditions, correspond to quantum-corrected black holes in $D$
dimensions, rather than classical ones.}
\end{quotation}

This conjecture follows naturally from the AdS/CFT correspondence
adapted to AdS braneworlds. According to it, the {\it classical}
dynamics in the AdS$_{D+1}$ bulk encodes the {\it quantum}
dynamics of the dual $D$-dimensional conformal field theory (CFT),
in the planar limit of a large $N$ expansion. Cutting the bulk
with a brane introduces a normalizable $D$-dimensional graviton
mode \cite{rs,addk}, while on the dual side this same
$D$-dimensional gravity mode is merely added to the CFT, which is
also cutoff in the ultraviolet. Then, solving the classical
$D+1$-dimensional equations in the bulk is equivalent to solving
the $D$-dimensional Einstein equations $G_{\mu\nu}= 8\pi
G_D\left<T_{\mu\nu}\right>_{CFT}$, where the CFT stress-energy
tensor incorporates the quantum effects of all planar diagrams. These
include particle production in the presence of a black hole,
and possibly other vacuum polarization effects.

This conjecture has implications in two directions.  On the one
hand, it allows us to view the brane-induced modifications of the
metric of a $D$-dimensional black hole as quantum corrections from
a CFT, a dual view that sheds light on both problems.  On the
other hand, we can use the conjecture to infer, from the known
properties of the classical bulk solutions, the properties
of the cutoff CFT coupled to gravity. Even if some of the
conclusions are derived using the AdS/CFT correspondence, they are
typically independent of the existence of a bulk dual: any
strongly coupled CFT with a large number of degrees of freedom is
likely to behave, when coupled to weak gravity, in a similar
manner.

We submit the conjecture to the test by reinterpreting the exact
solutions on the 2-brane in an AdS$_4$ braneworld \cite{ehm1,ehm2}
as quantum-corrected, gravitating CFT states in the dual $2+1$
theory, either with or without a negative cosmological constant in
2+1 dimensions, $\Lambda_3$. As is typical in tests of the AdS/CFT
correspondence, the calculations on the CFT side can only be
performed at weak 't~Hooft coupling, often at the one-loop order
only, and therefore comparisons with the strongly coupled dual of
the classical bulk theory, which includes all planar diagrams, are
difficult. Even then, we find some instances where the equivalence
between the results at weak and strong coupling holds to a great
degree of detail.

An interesting spin-off of the analysis is a realization of {\it
quantum censorship of conical singularities}, which we argue is a
generic effect independent of the AdS/CFT duality. Gravity in 2+1
dimensions is known to describe massive particles in terms of conical
singularities \cite{djt}. We find that when quantum corrections from a
CFT are included, the singularity of a sufficiently massive particle is
dressed by a regular horizon. This result is in fact true independently
of whether the CFT is strongly or weakly coupled, and acts more
efficiently when it has a large number of degrees of freedom.

Since we have a detailed description of the solutions in the
AdS$_4$ braneworld, we can apply it to describe the objects which arise in
the cutoff CFT. When $\Lambda_3 = 0$, the theory is characterized
by three mass scales: the UV cutoff of the CFT, $\mu_{UV}$, the 4D
Planck mass and the 3D Planck mass, in ascending order. These
scales naturally organize the range of CFT configurations
into three categories:
{\it (i)} the familiar light CFT states, with masses below the CFT
cutoff, which {\it are not} black holes because of the quantum
uncertainty-induced smearing; {\it (ii)} states with masses
between the CFT cutoff and the 4D Planck mass, which also {\it are
not} black holes because of quantum smearing and may receive large
quantum corrections in the bulk; and {\it (iii)} black holes,
which are the states with masses above the 4D Planck mass. These
black holes may be smaller than the CFT length cutoff,
$\hbar/\mu_{UV}$, but their description should be reliable since
both the bulk and the $2+1$ gravity corrections are small. Our
argument that the cutoff CFT can be trusted to distances much
shorter than the UV cutoff is analogous to a familiar situation in
string theory \cite{steve}, suggesting that the intermediate mass
states and light black holes behave as CFT solitons.

A negative cosmological constant $\Lambda_3 < 0$, allows for classical
BTZ black holes \cite{btz}. Although the AdS/CFT duality is not fully
understood for the case of negatively curved branes, we find that the
solutions localized on the 2-brane are naturally interpreted as BTZ
black holes with CFT quantum corrections, which are in equilibrium with
a thermal bath in AdS$_3$.  There are other localized solutions, all
with mass less than $M_{max} = 1/(24G_3)$, with different features, but
we find explanations for all of them within the context of our
conjecture. Black holes of mass larger than $M_{max}$ are delocalized
black strings occupying an infinite region of the bulk, and it is
unclear how to describe them within the confines of the $2+1$ theory;
in fact, it is likely that such a description should not be possible in
terms of only local physics.

In the physically more relevant case of a 3-brane in AdS$_5$ we
can not go into a similar level of detail
since there are no exact solutions, and
classical gravity in $3+1$ dimensions is dynamical.
However we can still explore the consequences of our conjecture in a
semi-quantitative manner. The description in terms of a CFT coupled to
gravity is not reliable until the horizon is larger than the
ultraviolet cutoff of the CFT, i.e., the black hole is sufficiently
heavy. For these black holes, the CFT+gravity theory allows us to
reinterpret the alleged obstruction for finding a static black hole
\cite{bgm} as a manifestation of the backreaction from Hawking effects.
The analysis of the trace anomaly of the CFT stress tensor allows us to
make this point precise. As long as the anomaly is consistent with the
asymptotic AdS$_5$ geometry, the conformal symmetry of the dual CFT is
valid in the infrared, and so there is no mass gap.
Hence any black hole at a finite temperature will emit
CFT modes as a thermal spectrum of Hawking radiation, which on the
bulk side is captured by a deformation of the bulk geometry
close to the brane, caused by the black hole sourcing the
classical gravity equations. We illustrate this to the leading
order on the CFT side by showing that the backreaction from Hawking
radiation, encoded in the form of a Vaidya-type far-field solution, is
consistent with the CFT anomaly. We also discuss the dual bulk picture
of Hawking radiation that arises from our conjecture. Within this
interpretation, the difficulties encountered in the ongoing quest for
the black hole localized on the 3-brane in AdS$_5$ are viewed as a
natural, subleading quantum correction to the classical solution,
rather than as a no-go theorem for
the existence of classical braneworld black holes.

\section{AdS/CFT duality for AdS Braneworlds}
\label{duality}

We begin with a brief review of several aspects
of the two dual descriptions that are relevant for our conjecture
\cite{malda}-\cite{apr7}.
Since we want to discriminate between classical and
quantum effects, we retain $\hbar$ in our formulas, while setting
$c=1$. Then, the $D$-dimensional Newton's constant $G_{D}$,
Planck length $\ell_D$, and Planck mass $M_D$ are related to each other as
\be
\label{newtons}
G_D = \frac{\ell_D^{D-3}}{M_D},\qquad \qquad \qquad M_D = \frac{\hbar}{\ell_D}\,.
\ee
In  AdS braneworlds the $D+1$ dimensional bulk Newton's constant
and the bulk cosmological constant $\Lambda_D=-D(D-1)/2L^2$ together
determine the Newton's constant induced on the $D$-dimensional brane
as
\be
\label{ans}
{G_D = {D-2\over 2 L}\, G_{D+1}\ .}
\ee
The precise details of the dual CFT depend on the specifics of the
string/M-theory construction that yield the AdS background. Here we
only need to know the effective number of degrees of freedom of the CFT,
$g_*$. For $D=4$, the dual pair are IIB string theory on
AdS$_5\times S^5$ of radius $L \sim \ell_{10}(g_s N)^{1/4}$
and ${\cal N}=4$ $SU(N)$ super Yang-Mills theory,
while for $D=3$, the dual pair are
M-theory on AdS$_4\times S^7$ and the (poorly known) theory describing
the worldvolume dynamics of a large number $N$ of M2 branes. In these
cases
\ba
\label{gstar}
&&g_* \sim N^2\sim
\left(L\over\ell_5\right)^3\sim \left(L\over\ell_4\right)^2
\quad (D=4), \nonumber\\
&&g_* \sim  N^{3/2} \sim \left(L\over\ell_4\right)^2
\sim {L\over \ell_3} \quad ~~~
(D=3) \, ,
\ea where we have used (\ref{ans}) to get the final expressions. $g_*$
is taken to be a large number, in order to keep small the quantum
corrections to the supergravity approximation to string/M-theory. For
the CFT, this is a large $N$ limit where planar diagrams give the
leading contribution.

The introduction of the brane that cuts off the AdS bulk implies that
very high energy
states of the dual CFT are integrated out, and the conformal invariance
of the theory is broken in the ultraviolet. However, the breaking
washes into the low energy theory only through irrelevant operators,
generated by integrating out the heavy CFT states at the scale
$\mu_{UV} \sim \hbar/L$. In the infrared, at energies $E<\mu_{UV}$, the
effects of the conformal symmetry breaking are suppressed by powers of
$E/\mu_{UV}$. Cutting off the bulk yields also a normalizable graviton
zero mode localized on the brane; this same $D$-dimensional gravity
mode is added to the dual theory. However, note that the CFT cutoff
$\mu_{UV}$ is {\it not} equal to the induced $D$-dimensional Planck
mass. Instead,
\be
\label{uvcutoff}
\mu_{UV} \sim
{M_4\over\sqrt{g_*}}
\quad (D=4), \qquad \qquad
\mu_{UV} \sim {M_3\over g_*} \quad
(D=3)\,,
\ee
which is much smaller than the Planck mass on the brane. The formulae
above can be written for any AdS space and can be viewed as a
definition of a cutoff CFT, although they do not guarantee the existence
of its UV completion. We will use them bearing this in mind.

\section{Quantum Black Holes on flat branes
in $2+1$ Dimensions}

For the case of $D=3$, the exact four-dimensional solutions
constructed in \cite{ehm1} yield the following
metric on the 2-brane,
\be
\label{bhbrane}
{ ds_{brane}^2 = -\left(1-{r_0 \over r}\right) dt^2
+ \left(1-{r_0 \over r}\right)^{-1} dr^2 + r^2 d\varphi^2} \, .
\ee
The parameter $r_0$ fixes the position of the horizon, and is determined
by the mass $M$. In a locally asymptotically flat space in $2+1$ the
mass is given by the conical deficit angle at infinity, $\delta_\infty=
8\pi\,G_3 M=8\pi\, M/M_3$. It was shown in \cite{ehm1} that such a
deficit angle is indeed present in (\ref{bhbrane}), leading
to\footnote{In the notation of \cite{ehm1}, $M_3$ was the mass as
measured on the brane, and $M_4$ the mass measured in the bulk. They
were shown to be the same, $M_3=M_4$. Here we denote them by $M$,
reserving $M_3$ and $M_4$ for the three- and four-dimensional Planck
masses,
as in eq.~(\ref{newtons}).}
\be
\label{masses}
M = \frac{M_3}{4} ~\Biggl(1 -
\frac{\sqrt{1+x}}{1+\frac32 x} \Biggr) \, ,
\ee
where $x$ is defined by
\be
\label{xm}
x^2(1+x) = \frac{r_0^2}{L^2}\,.
\ee
These expressions define the horizon size $r_0$ as a function of
the mass $M$ in parametric form.
The mass varies from $M = 0$ ($r_0=0$) up to a maximum,
\beq
M_{max} = 1/4G_3=M_3/4\,,
\eeq
which comes from the constraint that the deficit
angle $\delta_\infty$ be smaller than $2\pi$.
For small masses $M \ll M_3$
\be
\label{smallm}
r_0 \simeq \frac{4M}{M_3} ~L \ll L \, ,
\ee
while for the masses near $M_{max}$
\beq
\label{r0large}
r_0\simeq \frac{8L}{27 \left(1-M/M_{max}\right)^{3}}
\gg L\,.
\eeq

The presence of the horizon at $r=r_0$ may
appear as a surprise since it is known that there are no
asymptotically flat vacuum black holes in $2+1$ dimensions \cite{djt}.
But (\ref{bhbrane}) is not a vacuum solution. Following our conjecture,
it must admit an interpretation as a quantum-corrected solution of the
$2+1$ CFT+gravity system. To see this, note that the general relation
between the horizon radius and the mass is of the form
$r_0=L\,f(G_3M)$, with $f(G_3M)$ obtained from (\ref{masses}) and
(\ref{xm}). In order to correctly identify quantum-mechanical effects
we express the results in terms of only those variables which are
meaningful in the dual CFT+gravity description. Using (\ref{newtons}),
(\ref{ans}) and (\ref{gstar}) we can write $L\sim \hbar g_* G_3$, so
\be
\label{quanthor}
r_0\sim \hbar g_*G_3\, f(G_3M) \, .
\ee
The appearance of $\hbar$ is a clear fingerprint of the quantum origin
of the horizon viewed from the $2+1$ perspective. This is in complete
agreement with our conjecture: since there are no horizons in the
classical $2+1$ theory, any that are found must be purely
quantum-mechanical in origin. The classical theory does not contain any
length scale ($G_3M$ is dimensionless), and only with the introduction
of $\hbar$ can we form one, namely the Planck length $\ell_3=\hbar
G_3$, which sets the scale for $r_0$.

We can test the conjecture in more detail. The solution
(\ref{bhbrane}) can be formally obtained in the dual $2+1$ CFT coupled
to gravity from the quantum-mechanical backreaction on the spacetime of
a particle of mass $M$. Beginning with the conical geometry
corresponding to a localized CFT lump representing a point particle,
with deficit angle $\delta_{\infty} = 8\pi M/M_3$, one can compute the
Casimir stress-energy and find its backreaction on the metric. Such a
solution was indeed discovered almost a decade ago in \cite{soleng} for
the case of a weakly coupled scalar CFT. Its Casimir
stress-energy was computed in \cite{sas} as
\be
\label{stressmode}
\langle T^{\mu}{}_{\nu} \rangle= \frac{\hbar\alpha(M)}{r^3}\;{\rm
diag}(1,1,-2)\,,
\ee
where
\be
\label{alpha}
\alpha(M) = \frac{1}{128 \pi} \int^\infty_0 \frac{du}{\sinh u}
\left(\frac{\cosh u}{\sinh^3 u} - \frac{1}{(1-4G_3 M)^3}
\frac{\cosh[u/(1-4G_3 M)]}{\sinh^3 [u/(1-4G_3 M)]} \right) \, .
\ee
Using this stress-energy tensor to calculate the backreaction on the conical
spacetime, ref.~\cite{soleng} found the metric (\ref{bhbrane}), with
$r_0 =4\pi \hbar ~\alpha(M)/M_3 $. In our case the CFT has a large
number of degrees of freedom $g_*$, each of whom contributes to the
Casimir stress-energy tensor. Thus we expect to find $r_0 ={\cal O}(1)
~ \hbar g_* \alpha(M)/M_3$ where the ${\cal O}(1)$ factors can only be
calculated when the exact description of the strongly-coupled CFT is
known. Moreover, we can not expect the mass dependence of this $r_0$ to
agree precisely with that of (\ref{xm}) --- among other things, we have
not even included the contribution from fermions to $\langle
T^{\mu}{}_{\nu} \rangle$. Nevertheless, we may hope for some
simplification in the limiting cases $M \ll M_3$ and $M \rightarrow
M_3/4$. In the former limit,
\be
\label{alsmall}
\alpha(M) = {\cal O}(1) \frac{M}{M_3}  \, ,
\ee
so
\be
r_0 = {\cal O}(1) \hbar g_* \frac{M}{M^2_3} = {\cal O}(1)
\frac{M}{M_3} L\,,
\ee
which exactly reproduces eq.~(\ref{smallm}) up to ${\cal O}(1)$
coefficients. In the limit $M \rightarrow M_{3}/4$, the integrand in
(\ref{alpha}) is strongly peaked at $u=0$ and $\alpha(M)$ can be
computed using the saddle-point method,
\be
\label{saddle}
\alpha(M)= \frac{{\cal O}(1)}{ (1-4G_3M)^3}\, ,
\ee
so the backreaction from the CFT results in
\be
r_0 = {\cal O}(1)  \frac{\hbar g_*}{M_3(1-4G_3M)^3}
= {\cal O}(1) \frac{L}{(1-M/M_{max})^3}\,,
\ee
which again reproduces the precise parametric dependence in
eq.~(\ref{r0large}).

Alternatively, one can compare (\ref{stressmode})
with the stress-energy tensor computed directly from the metric
(\ref{bhbrane}),
\be
\label{stressehm}
T^{\mu}{}_{\nu} ={1\over 16\pi G_3}{r_0\over r^3}\;{\rm diag}(1,1,-2).
\ee
Both (\ref{stressmode}) and (\ref{stressehm}) have the same structure
and radius dependence, so they determine the same geometry. The
equivalence is completed by noting that, taking $g_*$ times
(\ref{stressmode}), and comparing to (\ref{stressehm}), we find $\hbar
g_* \alpha \sim r_0/G_3$, as expected. This formally confirms the
equivalence between the classical construction in AdS$_4$ and the
quantum-corrected $2+1$ solution.  The
quantum corrections are completely due to
Casimir-like vacuum polarization, rather than backreaction from Hawking
radiation, since the classical solutions are not black holes to begin
with. The Casimir effect acts here as a quantum censor, hiding the
classical conical singularity behind a horizon.

The agreement between the calculations in the two sides of the
conjecture is striking, given their completely different nature
(classical vs.\ quantum), and we
believe that it provides a strong argument in favor of the AdS/CFT
correspondence in the context of AdS braneworlds,
beyond the linearized calculation of \cite{apr5}.
One may ask whether the agreement is just a consequence of some common
symmetry underlying  both problems. This does not seem to be the case.
Conformal invariance is present on both sides: since the bulk AdS is
empty, it influences the brane only through the conformal Weyl tensor.
However, conformal symmetry alone only determines the radial
dependence $r^{-3}$ of the stress tensor (recall that the classical
$2+1$ theory has no length scale), and its traceless character.
Neither the particular structure ${\rm diag} (1,1,-2)$, nor the
dependence on the dimensionless quantity $M/M_3$, are fixed
by conformal invariance.

So far we have been focusing on the mathematical side of our conjecture
and ignoring the interpretation of the solutions (\ref{bhbrane}).
However, since we have argued that the solutions (\ref{bhbrane}) are
quantum-mechanical in origin, we must ask to what extent the
description of a state of mass $M$ based on (\ref{bhbrane}) is
physically valid. In particular, in the limit of small masses the
curvature of the solution will be very large outside of the horizon,
indicating that higher-order curvature corrections will invalidate the
solution (\ref{bhbrane}) already in a region larger than the horizon
size.

To understand the physics of the solutions (\ref{bhbrane}), note that
the states of the CFT+gravity theory are defined by {\it three} scales:
the CFT cutoff $\mu_{UV} \sim \hbar/L$ on the low end, the 3D Planck
mass $M_3$ on the high end, and the 4D Planck mass $M_4$ in between.
While $M_4$ is an obvious scale from the bulk side, from the viewpoint
of the dual CFT coupled to $2+1$ gravity its presence is slightly
mysterious. There, $M_4$ emerges because of the large number of CFT
degrees of freedom, as $M_4 \sim M_3/\sqrt{g_*}$. Its importance can be
seen as follows. Any solution of a given mass $M$ is characterized by
two length scales: the horizon radius $r_0$ and the Compton wavelength
$\lambda_C = \hbar/M$. If $\lambda_C > r_0$, the solution cannot be a
black hole, because quantum effects smear it over a volume larger
than the horizon, but if $r_0 > \lambda_C$, the solution is a black
hole, since quantum-mechanical fuzzying up is not sufficient to conceal
the horizon. On the bulk side, this simply means that the description
of this object by a {\it classical} metric in AdS space is not
appropriate, and that one should instead use wave packets delocalized
over $\lambda_C$ as in quantum mechanics. Viewed from the bulk it is
clear that the mass scale for the crossover is $M_4$. Translated into the
$2+1$ description, this is the same value at which $r_0\sim \lambda_C$:
when $M\sim M_4\ll M_3$,  (\ref{smallm}) and (\ref{ans}) imply $r_0\sim
L M_4/M_3\sim \hbar/M_4\sim \lambda_C$. Thus, $M_4$ is consistently the
threshold scale for black hole formation. Above this scale, the
curvature near the horizon is sub-Planckian, and the semiclassical
geometry (\ref{bhbrane}) becomes reliable all the way down to the black
hole horizon $r_0$.

Since for $M>M_4$ the leading CFT corrections are large enough to give
rise to a horizon, one may worry that higher order corrections may be
very large as well, and render the leading approximation meaningless.
Again, this does not occur. The higher-order effects in the $2+1$
description correspond to one-loop quantum effects (Hawking radiation)
in the bulk. The black hole temperature is $T\sim \hbar/r_0$, and when
$M > M_4$, $\hbar/r_0 \sim M^2_4/M < M$. Hence the backreaction will be
small, and the larger the horizon generated at the leading order, the
smaller the higher-order corrections outside it.

We stress that the quantum dressing of the conical singularity is in
fact completely independent of the AdS/CFT correspondence. It happens
for {\it any} $2+1$ CFT that couples to $2+1$ gravity, independently of
whether its ('t~Hooft) self-coupling is strong or weak. While
ref.~\cite{soleng} claimed that when $g_* = 1$ the solutions
(\ref{bhbrane}) are never reliable, because of large quantum
corrections outside of the horizon, this is true only in the regime of
small masses. In the limit $M \rightarrow M_{max}$ the horizon becomes
arbitrarily large, (\ref{r0large}), and the solution (\ref{bhbrane})
{\it is} a black hole. The main feature here is that the regime of
intermediate mass states disappears as $g_* \rightarrow 1$ because
$\mu_{UV} \rightarrow M_4 \sim M_3$, and the transition between light
states and black holes is sudden. Adding a large number of degrees of
freedom expands down to $M_3/\sqrt{g_*}$ the range of masses where the
horizons can be trusted and makes quantum cosmic censorhip more
efficient. Note that
these quantum corrected black holes have a large entropy  ($\propto$
the area in the bulk, not on the brane \cite{ehm1}), and that at first
sight its origin may be puzzling, considering the fact that the
classical background which gave rise to this was modeled as a cone
sourced by a point-like distribution of CFT energy. However, this
source should really be viewed not as an individual state but as a lump
of many CFT degrees of freedom, whose entropy is resolved with the
help of gravity and quantum corrections.

Therefore the CFT objects fall into three classes as a function of their
mass:

{\bf 1)} Light states with masses $M < \mu_{UV}$ with $\lambda_C \gg
r_0$, and so they cannot be reliably described by (\ref{bhbrane}). They
require a quantum-mechanical description in the bulk independently of
the localized $2+1$ gravity, and on the AdS$_4$ side are just the
perturbative massive KK modes \cite{rs}.

{\bf 2)} Intermediate mass objects $ \mu_{UV} < M < M_4$, with $\lambda_C
> r_0$, and so they too are not black holes. Since their masses are
above the cutoff, they cannot be described as bulk KK modes on the
AdS$_4$ side. They are new nonperturbative states, which are bulk
deformations of AdS$_4$. Their detailed properties are sensitive to the
physics at the cutoff scale. If the only new mode which appears at the
cutoff is $2+1$ gravity (a non-dynamical mode), they can be viewed as
bound CFT states, which may however receive large bulk quantum
corrections that are not automatically under control because $\lambda_C /r_0 > 1$.

{\bf 3)} Heavy objects $M_4 < M \le M_{max}$ with $\lambda_C < r_0$,
and so they really are black holes. As with the intermediate mass
states, the description of the black holes with $M_4 < M \ll M_3$
requires physics at distances shorter than the CFT cutoff $L$, which
may be completely reliable if the only new mode at the cutoff is the
$2+1$ gravity. Then both the $2+1$ corrections from the graviton and
the bulk quantum corrections remain small since they are proportional
to $T/M = \hbar/r_0 M < 1$, as seen above. These black holes
are unstable to the emission of Hawking radiation, which on the bulk
side is a one-loop effect, corresponding to non-planar diagrams in the
CFT dual.

The emergence of the new short distance scale $\ell_4 = \hbar/M_4\ll L$
is analogous to the emergence of very short distance scales $\ell_* =
g_S \ell_S$ in string theory, which can be probed by solitonic objects
- the $D$-branes \cite{steve}.

In closing, we define how to take the classical limit
for the $2+1$ theory in a way in which the black holes survive. To
identify the appropriate limit, observe from (\ref{quanthor}) that to
keep the horizon finite we must take simultaneously $\hbar \rightarrow
0$ and $g_* \rightarrow \infty$, with $\hbar g_*$ finite. Since also
$L=\hbar g_*\,G_3$ and $G_4 = \hbar/M^2_4 \sim L G_3$ stay finite, the
bulk description remains valid.  Consider now the black hole entropy $S
= \pi g_*\, x^2/(2+3 x)$ and the temperature $T = \mu_{UV} /[4\pi x
\sqrt{1+ x}]$. Since $x$ is a function of only $G_3M$ through
(\ref{masses}), $S$ and $T$ are written in terms of $2+1$ quantities
only. Both are formally independent of $\hbar$, and naively seem to
remain constant as $\hbar \rightarrow 0$. However, taking also
$g_*\to\infty$, the black hole temperature vanishes and its entropy
diverges, as they should.

\section{Quantum Black Holes in $2+1$ Dimensions with $\Lambda_3 < 0$}

Due to the peculiarities of $2+1$ gravity, in the previous example the
black hole horizon arises only after the leading quantum corrections
are included. Hawking radiation and its backreaction will not appear
until the next order, which is difficult to compute. By contrast,
classical gravity in $2+1$ dimensions with a negative cosmological
constant admits not only the conical spacetimes of point particles, but
also classical (BTZ) black holes \cite{btz}. Spacetimes with a negative
cosmological constant can also be constructed as AdS bulk geometries ending
on negatively curved branes if their tension does not satisfy the RS2
fine-tuning \cite{nk}. Black holes on negatively curved 2-branes in
AdS$_4$ have been constructed in \cite{ehm2}, so we can use these
solutions to study further our conjecture.

However, the bulk geometry at large distances from negatively curved
branes differs in important ways from the bulk surrounding the flat
branes discussed previously. The proper size of radial slices decreases
away from the brane until a minimal size, a throat, is reached, after
which the space re-expands again. Therefore the total bulk volume is
infinite. Because of this, the solutions with horizons can be either
black holes localized on the brane, or black strings stretching all the
way through the AdS space, depending on their mass. A second,
positive tension, regulator brane may or may not be introduced to cut
this volume off. If the regulator is included, then the relationship
between $G_3$ and $G_4$ changes to
\cite{ehm2}
\be
\label{newtbtz}
G_3=\frac{1}{2\sqrt{\lambda}\,L_3}G_4\, ,
\ee
where $L_3$ is the length scale of the brane cosmological constant,
$\Lambda_3=-1/L_3^2$, and $\lambda$ is a dimensionless parameter defined
by
\be
\lambda\equiv{L^2\over L_3^2-L^2}\, .
\ee
If the brane is only slightly curved, $L_3\gg L$, i.e., $\lambda\simeq
L^2/L_3^2\ll 1$, we recover (\ref{ans}) approximately. The duality as
described in Sec.~\ref{duality} can not be applied in a straightforward
manner: the holographic dual is modified in the infrared, and is
considerably less understood than in the case of flat branes
\cite{kr,br,porrati}. Essentially, in this case the presence of the brane
that breaks conformal symmetry in the UV communicates the breaking to the
IR as well. This can be easily seen on the bulk side. Consider the setup
with a regulator brane on the other side of the throat. This ensures
the validity of $2+1$ gravity at all length scales, but it alters the
CFT in the IR by introducing an IR cutoff. The CFT
states fall into a discrete spectrum,
with a mass gap that scales as the IR cutoff, $\mu_{IR}
\sim \hbar/L_3$. In the limit when the regulator is removed, the gap
does not disappear: the fluctuating bulk modes, which correspond to the
CFT states, must obey Dirichlet boundary conditions at the AdS
boundary to remain normalizable. Thus the presence of the AdS brane
leads to a two-sided boundary value problem and the
spectrum remains quantized.

The mass gap suppresses Hawking emission for very cold, small black
holes, because their temperature is below the gap and so the CFT modes
cannot be emitted as thermal radiation. Then, to leading order the
backreaction for these would be very suppressed as long as the
temperature is below the gap. Other consequences of the mass gap will
be apparent near the end of this section. In the following we will work
in the approximation where $\lambda$ is small, so the IR and UV
regulators are well separated and (\ref{ans}) remains approximately
valid.

Besides Hawking emission, we expect quantum corrections from the
Casimir effect induced, as in the previous section, by the
identifications of points in the background. In the cases where the
horizon is absent (or has zero temperature) at the classical level, the
thermal Hawking radiation will be absent. But for a BTZ black hole, it
is difficult to distinguish between thermal and Casimir effects.
Actually, the distinction is rather artificial, since both arise from
the same non-trivial identifications of points in AdS$_3$.

We begin the analysis with the solution for a localized black hole on a
negatively curved 2-brane found in \cite{ehm2},
\be
\label{btzbrane}
{ ds_{brane}^2 = -\left(\frac{r^2}{L_3^2}-8G_3 M -\frac{r_1(M)}{r} \right)
dt^2 + \left(\frac{r^2}{L_3^2}-8G_3 M -\frac{r_1(M)}{r} \right)^{-1} dr^2 +
r^2 d\varphi^2} \, ,
\ee
which is asymptotic to AdS$_3$. This is similar to the BTZ black hole of
mass $M$, with an extra term $r_1(M)/r$. As in the previous example,
$r_1(M)$ can only be given in parametric form. Defining a parameter $z$
via
\be
G_3M=\frac{z^2(1+z)(\lambda-z^3)}{2(\lambda +3z^2+2z^3)^2} \, ,
\ee
then
\be
r_1=8L_3\sqrt{\lambda}\;
\frac{z^4(\lambda+z^2)(1+z)^2}{(\lambda +3z^2+2z^3)^3}\, .
\ee

\FIGURE{ \epsfig{file=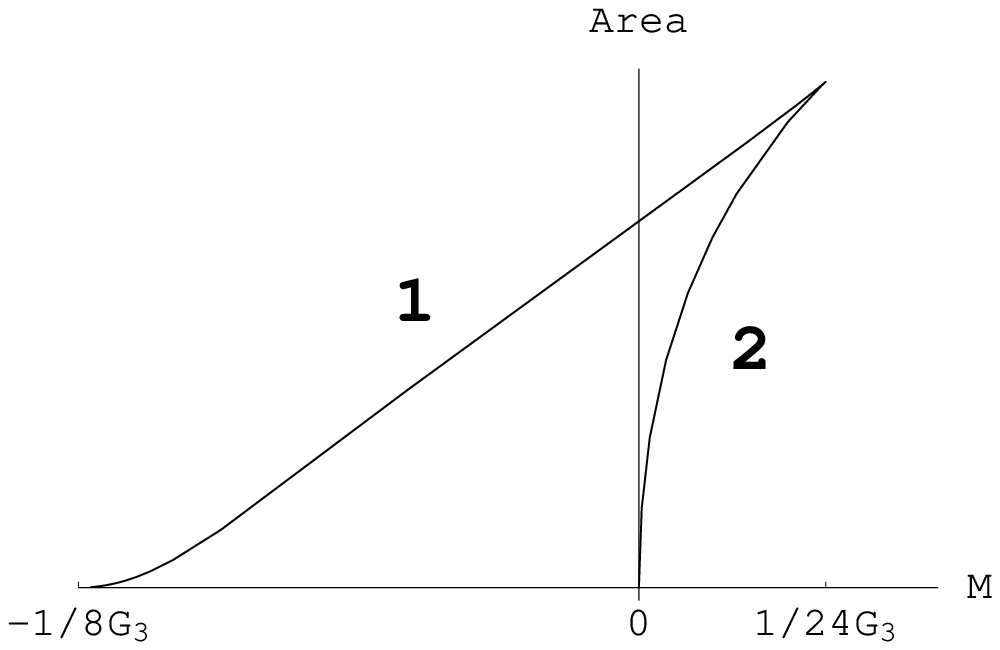, width=6.5cm, height=6cm}
\caption{Mass dependence of the 4D area of
black holes on an AdS$_3$ brane.} \label{fig-coordinates}}

The range of masses in (\ref{btzbrane}) which do not lead to naked
singularities or to delocalization of the black hole into a black
string is $-1/8G_3\leq M\leq 1/24G_3$ (obtained by varying $z\in
[0,\infty)$). For $M=M_{min}=-1/8G_3$ the correction term
vanishes, $r_1=0$, and one recovers AdS$_3$ in global coordinates.
The range $-1/8G_3< M<0$ corresponds, in classical vacuum gravity,
to conical singularities, but here they are dressed with regular
horizons. In Fig.~1 we display the bulk horizon area of all these
solutions \cite{ehm2}. This helps us identify two branches of
solutions: the branch labeled {\bf 1} starts at $M=-1/8G_3$ and
ends at $M= 1/24G_3$. Branch {\bf 2} begins at $M=0$ and zero
area, and ends at the same point as the previous one.

As before, (\ref{btzbrane})
does not solve the vacuum Einstein equations
with a negative cosmological constant. Instead, the stress-energy
tensor that supports (\ref{btzbrane}) contains a correction
of the form
\be
\label{btztensor}
T^{\mu}{}_{\nu}=\frac{1}{16\pi G_3}\frac{r_1(M)}{r^3}
\;{\rm diag}(1,1,-2)\, .
\ee
We must discuss how, in accord with our conjecture, these terms encode
the quantum effects in the dual theory.

The sector $-1/8G_3\leq M<0$ of the first branch is naturally
interpreted as in the previous section: these solutions are
classical conical spacetimes dressed with a horizon from the
backreaction of the Casimir energy of the CFT.
We are not aware of any calculations of the Casimir energy of a
conformal field in conical ($M<0$) AdS$_3$ spacetimes, nor of its
backreaction. However, we can verify the correspondence between this
sector and the one of the previous section, in the limit where the
cosmological constant vanishes, $L_3\to\infty$. If we take this limit
for the solutions (\ref{btzbrane}) and rescale the time and radial
variables to their canonical form at infinity, we find
\be
\label{flatlimit}
{ ds_{brane}^2 \to -\left(1-{r_1 \over (8G_3|M|)^{3/2}r}\right) dt^2 +
\left(1-{r_1 \over (8G_3|M|)^{3/2}r}\right)^{-1} dr^2 + 8G_3|M|r^2
d\varphi^2} \,.
\ee
This has the same form as (\ref{bhbrane}), with $r_0$ identified as
$r_1/(8G_3|M|)^{3/2}$. The mass of the limiting solution, $\tilde M$,
obtained from the conical deficit in (\ref{flatlimit}), is
\be
\tilde M=\frac{1}{4G_3}\left(1-\sqrt{8G_3|M|}\right)\,.
\ee
The masses in asymptotically flat and AdS spaces are differently
measured, so it is not surprising that $\tilde M$ differs from $M$.
What is important is that the range of masses $-1/8G_3\leq M \leq 0$
maps precisely to the range in asymptotically flat space, $0\leq \tilde
M \leq 1/4G_3$.  One can also check that in the limit $L_3\to\infty$,
$r_1/(8G_3|M|)^{3/2}$ as a function of $\tilde M$ becomes exactly the
same as $r_0$ in (\ref{masses}) and (\ref{xm}), with the identification
$z\to 1/x$. Hence we are quite confident that this sector of AdS$_3$
solutions can be interpreted as Casimir-censored singularities, and where
the censorship is reliable for sufficiently large masses $\tilde M$,
as before.

In the sector $0\leq M\leq 1/24G_3$ there are two branches of
black holes. For a given mass, branch 1 solutions have larger
area than branch 2. We will see that the interpretation
is clearer for the solutions in branch 1.

For a conformally coupled scalar at weak coupling
residing in the BTZ background, the renormalized stress tensor $\langle
T^{\mu}{}_{\nu}\rangle$ has been calculated in \cite{steif,lo,shm}, and
it has the same structure as (\ref{stressmode})\footnote{The form of
$\langle T^{\mu}{}_{\nu}\rangle$ depends on the boundary conditions at
the AdS$_3$ boundary. The brane solutions appear to automatically select
`transparent' boundary conditions, while ref.~\cite{lo} considers
instead Dirichlet or Neumann conditions. The results for transparent
conditions follow by omitting all terms in \cite{lo} with a ``$\pm$''
factor, bringing \cite{steif,lo,shm} into agreement.},  now with
\be
\label{alphabtz}
\alpha(M)=\frac{(8G_3M)^{3/2}}{16\sqrt{2}\pi}
\sum_{n=1}^\infty \frac{\cosh 2n\pi\sqrt{8G_3M}+3}
{\left(\cosh 2n\pi\sqrt{8G_3M}-1\right)^{3/2}}\,.
\ee
Since this $\langle T^{\mu}{}_{\nu} \rangle$ has the same structure as
the brane stress-energy tensor (\ref{btztensor}), the backreaction
calculated in \cite{lo,shm} results in a geometry like the brane metric
(\ref{btzbrane}).

This stress-energy tensor is not of the thermal type $\propto {\rm
diag} (-2,1,1)$. However, this does not conflict with the fact that the
CFT in the presence of the black hole is in a thermal state.
Ref.~\cite{lo} showed that the Green's function from which this
$\langle T^\mu{}_\nu\rangle$ is derived is periodic in imaginary time,
with a period equal to the local Tolman temperature dictated by the
black hole. Moreover, this Green's function satisfies the analyticity
properties that characterize the Hartle-Hawking state. This means that
there is a thermal component in the stress-energy tensor of the CFT, in
static equilibrium with the black hole. The fact that the tensor
structure of $\langle T^\mu{}_\nu\rangle$ does not conform to the
canonical thermal one near infinity reflects the presence of a large
Casimir contribution.

For the $M=0$ black hole, which has zero temperature in the classical
limit, one would expect that the backreaction from Hawking radiation is
absent at one loop. In this limit \be
\alpha(0)=\frac{\zeta(3)}{16\pi^{4}}\,, \ee which is finite and
parametrically ${\cal O}(1)$, i.e., not small. This indicates that the
quantum-corrected solution undergoes a large Casimir backreaction and
cannot be the massless zero-area solution in the second branch, but
rather the black hole in the first branch, of finite size. For this
state \be r_1(0)=\frac{8}{27}L={\cal O}(1)\;\hbar g_* G_3\,, \ee i.e.,
$r_1(0)/G_3={\cal O}(1)\; \hbar g_* \alpha(0)$, and so the brane and
CFT stress-energy tensors agree and the interpretation is consistent.
The same is true for all $M>0$ black holes in the first branch: the
dependence of $r_1$ on $M$ is weak when $\lambda\ll 1$, so $r_1(M)$
remains $\sim L$, and similarly $\alpha(M)={\cal O}(1)$ in the range of
masses $0<M<1/24G_3$, so we find the same agreement up to numerical
factors. It is difficult to compare the mass dependence with the same
level of rigor as in the asymptotically flat case. For example, the
fermions are typically much more sensitive to the cosmological constant
than scalars, and so the details of the mass dependence of the function
$\alpha(M)$ for the complete dual CFT, even if we ignore the effects of
strong coupling, will be quite different from the scalar contribution
(\ref{alphabtz}). For the largest possible masses,
$M\approx 1/24G_3$, the temperature of the black hole is of the order
of the IR cutoff, $\sim \hbar/L_3$, and hence Hawking radiation is not
suppressed. One may say that it becomes comparable to the Casimir
energy, but it is difficult to tell one from the other.

Therefore, all branch 1 solutions at least fit consistently with our
conjecture. The black holes of branch 2 may also allow an
interpretation as follows. In our conjecture, no specification is made
of what is the vacuum state of the CFT. In particular, the calculation
of $\alpha(M)$ in \cite{steif,lo,shm} was performed assuming that the
state in which $\langle T_{\mu\nu}\rangle$ vanishes is the global
AdS$_3$ vacuum. However, it is also possible to regard the $M=0$ state
of zero area as a consistent vacuum, in which case the stress tensor
would be renormalized so that $\langle T_{\mu\nu}\rangle_{M=0}=0$. This
$M=0$ black hole would remain uncorrected, and the BTZ black holes with
backreaction from a CFT state above this vacuum would result in a
branch of solutions starting at zero area at $M=0$, just like branch 2.
While it is difficult to test this idea further, it is tempting to
speculate with the possibility of a decay of the $M=0$ vacuum by making
a transition to the more entropic $M=0$ state of branch 1, followed by
evaporation down to the global AdS$_3$ vacuum\footnote{In the presence
of supersymmetry, these two vacua differ in the periodicity conditions
for fermions, as NS or R vacua, and therefore fall into different
superselection sectors.}.

Finally, we comment on the solutions with masses $M>1/24G_3$, which
also exist when $\Lambda_3 < 0$. The metric they induce on the brane is
precisely BTZ without any corrections. In the bulk, these black holes
are in fact black strings that stretch beyond the throat region, all
the way to the AdS boundary on the other side. Therefore they are
extremely sensitive to the infrared modifications in the dual picture,
and their full dynamics is clearly not amenable to the description in
terms of only $2+1$ CFT+gravity theory. While the apparent absence of
quantum corrections to these black holes seems puzzling, a possible
resolution is that these black holes are so massive that the
backreaction on them is not only small, but even vanishing at the level
of planar diagrams. Note that the one-loop stress-energy tensor of the
CFT at weak coupling becomes exponentially small in $\sqrt{M}$ for
large $M$ (see (\ref{alphabtz})), which may be an indication of such
behavior. Another indication comes from the higher-dimensional nature
of these solutions: since they extend through the throat, these
solutions cannot be described by $2+1$ gravity. Instead, for them the
$2+1$ gravity effectively decouples, and their temperature should be
viewed as a purely bulk loop effect, with $G_3 M$ reinterpreted as $G_4
m$, where $m$ characterizes the mass per unit length of the string. We
postpone a detailed consideration of these solutions for future work.

\section{Resolving the Mystery of the Missing $3+1$ Black Hole}

We now turn to the Randall-Sundrum model \cite{rs}, defined by a single
3-brane in the AdS$_5$ bulk. We have far less control over the theory
now: on the one hand, gravity in $3+1$ dimensions is dynamical; on the
other hand, the absence of exact solutions makes the identification of
CFT states difficult. Let us proceed by analogy with the $2+1$
analysis. In that case black holes of horizon size $r_H=r_0<L$ are
approximately spherical four-dimensional black holes in the bulk. This
feature extends to higher dimensions. Quite generally, a black hole of
size $r_H$ on the brane has an extent into the bulk $r_B \sim L
\ln(1+r_H/L)$, so at distances $r_H < L$ the bulk solution becomes
progressively less flattened around the brane and rounder, $r_H\sim
r_B$. In the present context, it is well approximated, near the
horizon, by a five-dimensional Schwarzschild solution.  As $r_H$
becomes smaller than $L$ an increasing number of CFT modes in the UV
must be interpreted as bulk gravity in order to encode the bulk
geometry. Then it is not meaningful to describe the state as a
CFT-corrected 3+1 black hole. The situation in 2+1 dimensions was in
this regard better than one had any right to expect, since the picture
of a classical solution, the conical singularity, dressed by CFT
corrections was actually valid for masses all the way down to the scale
$M_3/\sqrt{g_*}\sim M_4$, i.e., distances {\it much smaller} than the
CFT length cutoff $\hbar/\mu_{UV}\sim L$. The reason is that pure
classical gravity in $2+1$ dimensions is topological, so the CFT
corrections give the leading dynamical effects of gravity.  In that
case, the length scale $r_0\sim L$ does not determine any
parametrically new mass scale.

Instead, in $3+1$ dimensions the transition point defined by the
equality $r_H\sim L\sim G_4 M\sim (G_5M)^{3/2}$ determines, through
(\ref{newtons}), (\ref{ans}) and (\ref{gstar}), the new mass scale
$\sqrt{g_*} M_4$. We can not sensibly describe black holes lighter than
this as CFT-corrected $3+1$ black holes. Nevertheless, the bulk
description holds as long as the backreaction in the bulk remains
small. This is the case if $M>M_5\sim M_4/g_*^{1/6}$. This suggests
that the small black holes above this scale are additional states of
the CFT, besides the light modes of mass $M<\mu_{UV}$. However, since
they are very sensitive to the UV regulator of the CFT, they are not
suitable for testing our conjecture. Only for $M>\sqrt{g_*} M_4$ can
the light bulk KK modes be consistently interpreted as modes of a CFT
and not as gravity.

Therefore, in what follows we will focus on black holes with mass $M >
\sqrt{g_*} M_4$, i.e., size $r_0>L$. Since their mass is much greater than
$M_4$, the backreaction of $\left< T_{\mu\nu} \right>$ can be regarded as a small
perturbation of the classical black hole solution and treated order by
order as an expansion in $\hbar$. In general, $\left< T_{\mu\nu}
\right>$ depends on the definition of the quantum vacuum in a crucial
way \cite{books,bfs}. There are three usual choices, each describing a
different physical situation:

{\bf (1)} The Hartle-Hawking state, which describes a black hole in a
thermal bath in equilibrium with its own radiation. The state of the
CFT is regular at the event horizon. Far from the black hole $\left<
T_{\mu\nu} \right> $ describes a gas of 4D CFT radiation at the Hawking
temperature. This is incompatible with asymptotic flatness. A natural
possibility is that a small backreaction results in an
FRW universe containing a black hole immersed in thermal radiation.

{\bf (2)} The Unruh state, which describes the process of black hole
evaporation. The stress-energy tensor is regular only at the future
horizon, and there is a thermal flux of radiation at future null
infinity. Consistent backreaction must produce a time-dependent,
quantum-corrected, evaporating black hole solution.

{\bf (3)} The Boulware state, which describes a static configuration,
with a stress-energy tensor that vanishes at infinity but diverges at
the horizon. The backreaction effects convert the horizon into a null
singularity. This singularity can be cut away by a static interior
solution if it is greater than the singular
surface, such as a star.

According to our conjecture, the solution for a black hole on the
RS2 brane must correspond to one of these choices. It is now
obvious why the search for a static, asymptotically flat black
hole solution on the brane has been fruitless so far: the state
(1) is not asymptotically flat, (2) is not static, and (3) does
not have a regular horizon. The physical reason why we expect that
the black hole should sense the backreaction is easy to see from
AdS/CFT. As long as the bulk has asymptotic AdS$_5$ geometry, on
the dual side the conformal symmetry of the CFT is valid in the
infrared, and so there is no mass gap separating the CFT modes from the
vacuum. Any black hole at a {\it finite} temperature will therefore emit
CFT modes with a thermal spectrum, which is precisely the Hawking
radiation\footnote{In the case of RS2 in AdS$_5$ a step towards the ideas
presented here was entertained by T.~Tanaka \cite{tanaka}, and,
simultaneously, by R.~Maartens and one of us (NK) in the discussions
reported in \cite{magdr}, in order to explain the results of \cite{bgm}.
A naive argument that the bulk dynamics encodes the backreaction
from Hawking radiation would lead one to expect that {\it all}
asymptotically flat brane-localized black holes are
time-dependent. This would be in conflict with the exact
static $2+1$ solutions of
\cite{ehm1,ehm2}. Our conjecture that the classical bulk dynamics
encodes {\it all} quantum corrections at the level of planar diagrams
completely resolves this conflict. These exact solutions in fact strongly
support the conjecture.}. On the bulk side, this must be described by a
deformation of the bulk geometry near the brane, which arises
because the black hole appears as a source in the classical bulk
gravity equations.

We should recall here some proposals for static black hole solutions on
the brane. For reasons that will become clearer later, such solutions
typically become singular in the bulk, so they are not physical. A
prototype for this sort of singular behavior is the black string of
\cite{chr}. Although the brane metric is perfectly regular, there is a
divergence of the curvature at the Cauchy horizon in the bulk.

The preceding discussion naturally leads us to considering a radiative
solution as the leading-order description of the exterior of a black
hole localized on the brane. The detailed description of this geometry
on the bulk side would require either the exact bulk solution, which
has been missing so far, or a much better approximation than the
existing ones. On the side of the $3+1$ CFT+gravity, a description at
the same level of rigor would require a careful backreaction analysis,
where we should start with a classical Schwarzschild black hole and
perturb it by means of the $\left< T_{\mu \nu}\right> $ in the Unruh
state evaluated in the classical background geometry.  This analysis
rapidly becomes quite involved, because of the necessity for describing
the near and far field regions of the black hole differently: a
negative energy density flux near the horizon, well approximated by an
ingoing Vaidya metric; the asymptotic infinity approximated by an
outgoing Vaidya metric, and a complicated geometry describing the
transition between these asymptotic forms in between. The far-field
outgoing metric encodes the flux of Hawking radiation pouring out of
the black hole, which is described by the stress-energy tensor
\be
\label{radiation}
T_{\mu\nu} = \frac{{\cal L}(u)}{4\pi r^2} \nabla_\mu u \nabla_\nu u \, ,
\ee
where $u$ is the retarded null coordinate and
${\cal L}(u)$ is the flux luminosity. The perturbed geometry is
\be
\label{radmetric} ds^2 = -\left(1-\frac{2G_4 M(u)}{r}\right) du^2
- 2 drdu + r^2 d\Omega_2 \, ,
\ee
where $\frac{dM(u)}{du} = - {\cal L}(u)$.  To check our conjecture, we
should recover the relation between ${\cal L}$ and $M$ from
leading-order corrections to the black hole geometry induced from the
bulk. To make any such calculation precise, we should relate the
far-field solution (\ref{radmetric}) to a near horizon one, and then match
this solution to the interior. The matching conditions will give
the precise form of the relationship between the luminosity ${\cal L}$
and the interior parameters.

In order to circumvent the details of the matching between the near and
far regions, we resort to a simpler, heuristic calculation that allows
us to reproduce the correct parametric dependence of the
luminosity. Consider the radiative collapse of a large dust cloud. Match
this collapsing cloud of dust, whose dynamics is determined in
\cite{bgm} by a leading-order bulk calculation, to an outgoing Vaidya
metric (\ref{radmetric}), following the work of
\cite{santos}\footnote{This appears in Ref.~\cite{dad}, who,  however,
had an outgoing Vaidya metric everywhere outside the collapsing sphere,
and also continued matching this solution to a Reissner-Nordstrom
geometry very far away. This latter step seems dubious, because this
geometry is very likely singular in the bulk.}. The quantum
correction terms propagate through the matching regions, and
this relates the outgoing flux of radiation to the {\it subleading}
correction in the interior star geometry, which is $\propto (G_4 M
L)^2/R^6$, as calculated in \cite{bgm}, r.h.s.\ of their eq.~(6) (we
only consider the limit $Q=\Lambda=0$ of this equation, which is
sufficient for our purposes). Comparing to (\ref{radiation}) we find
${\cal L} \sim G_4(M L)^2/R_0^4\sim \hbar g_* (G_4 M)^2/R^4_0$, where
$R_0$ is the radius of the matching surface. For a large collapsing
mass, this will be near $2G_4 M$, so ${\cal L} \sim
\hbar g_*/(G_4 M)^2$. This is the value that corresponds to a flux of
Hawking radiation of $\sim g_*$ degrees of freedom of the CFT, at a
temperature $T_H \sim \hbar/(G_4 M)$, as required. Replacing $M(u)$ by
$M$ is consistent since ${\cal L} \propto \hbar$ and we are working in
an expansion in $\hbar$. Within this approach we cannot
obtain a detailed formula with accurate numerical coefficients, but
it does reproduce the correct scalings with the black hole and CFT
parameters, in complete accord with our conjecture. A more detailed
analysis recovering the precise form of the matching conditions would
be useful, since it can display how the outgoing flux is turned on as a
function of time.

What remains is to verify the consistency of the matching of geometries
across the horizon. A simple way to check this is to compare the
quantum trace anomalies of the backreacted states in the exterior and
interior. The trace anomaly of the quantum stress tensor is a local
geometric quantity independent of which vacuum the field is in
\cite{duff,cf}. It has been studied in detail in the AdS/CFT context
\cite{skhe}, and in particular in the case of AdS braneworlds in
\cite{skender,shiroida,soda}. It gives us further insight into our problem,
in that it provides a simple leading-order consistency check, which a
configuration must pass in order to be described by the leading-order
effects in the duality pair.

For a weakly coupled CFT in $3+1$ dimensions
the trace anomaly $ \left< T^\mu{}_\mu \right> $
is, to leading order, \cite{books}
\be
\label{an} \left< T^\mu{}_\mu \right> =
\frac{\hbar}{(4\pi)^2}(aC^2+bE +
c\nabla^2 R),
\ee
where $C^2=R_{\mu\nu\alpha\beta}R^{\mu\nu\alpha\beta}
-2R_{\alpha\beta}R^{\alpha\beta}+R^2/3$ is the square of the Weyl
tensor,
$E=R_{\mu\nu\alpha\beta}R^{\mu\nu\alpha\beta}-
4R_{\alpha\beta}R^{\alpha\beta}
+R^2$ is the Gauss-Bonnet term. The coefficients $a$, $b$ and $c$
depend on the specific matter content of the theory, and
in the case of $D=4$ ${\cal N}=4$ $SU(N)$ SYM at large $N$
\be
\label{susyan}
\left< T^\mu{}_\mu \right> =
\frac{\hbar N^2}{32\pi^2}\left(R^{\mu\nu}R_{\mu\nu} -
\frac{R^2}{3} \right) \,.
\ee
Note the cancellation of the term
$R_{\mu\nu\alpha\beta}R^{\mu\nu\alpha\beta}$.  Ref.~\cite{skhe} showed
how this anomaly is precisely reproduced from a computation in the
AdS$_5$ bulk. This result is perturbatively identical to the familiar
quadratic stress-energy correction terms that appear in the effective
long distance $3+1$ gravity equations in AdS braneworlds \cite{mss},
which can be checked explicitly recalling $g_* \sim N^2$
\cite{skender,shiroida}.

If the CFT is deformed by relevant operators the behavior in the
infrared changes, and the bulk side of the geometry will be quickly
deformed away from the AdS geometry. When this occurs, the anomaly
coefficients $a,~b,~c$ in (\ref{an}) will deviate away from the values
they take for ${\cal N}=4$ SYM, and generically $a+b\neq 0$, so the
anomaly may contain the contributions from
$R_{\mu\nu\alpha\beta}R^{\mu\nu\alpha\beta}$. The appearance of such
terms implies that the bulk is {\it not} asymptotically AdS$_5$; it is
very likely that a singularity will appear in the bulk, at some finite
distance from the brane\footnote{The exception are the situations where
the singularity can be dealt with in a physically motivated manner. For
instance, a singularity appears when supersymmetry is broken to produce
either a confining phase or a mass gap at some finite scale in the
infrared, and its resolution is an interesting problem
\cite{polstra}.}. On the other hand, the absence of terms $\propto
R_{\mu\nu\alpha\beta}R^{\mu\nu\alpha\beta}$ does not imply that the
bulk is asymptotically AdS. An example is a radiation dominated FRW
cosmology, with the CFT in a thermal state. In the bulk, this
corresponds to an AdS-Schwarzschild solution, where the singularity is
hidden by a horizon at a finite distance from the brane
\cite{kraus,apr3,hhr,hmr}, although the anomaly vanishes.

We can now reinterpret the
`no-go theorem' of ref.~\cite{bgm} within the CFT+gravity theory.
There the authors considered the collapse of
pressureless homogeneous dust on a braneworld in AdS$_5$, and following the
standard general relativity routine, they attempted to match
this interior to an exterior metric, as opposed to a radiating
one as we advocated above. Because the interior geometry was a
solution of the AdS$_5$ braneworld junction conditions, it
was guaranteed to satisfy the anomaly equation (\ref{susyan}).
However, the exterior geometry, resembling a deformation of the
Schwarzschild geometry, was not required to fulfill these equations,
but was tailor-made to satisfy the matching conditions on the envelope
of the collapsing dust. Requiring the exterior geometry to be static,
ref.~\cite{bgm} found that the Einstein tensor must have a
nonvanishing trace in the exterior region equal to
\be
\label{nrs}
G^{\mu}{}_{\mu}=  -12 L^2\frac{(G_4 M)^2}{r^6}\,.
\ee
This led \cite{bgm} to conclude that the exterior geometry can not be
static.

The interpretation of this result
is that (\ref{nrs}) is the quantum anomaly induced by the backreaction,
which is inconsistent with the anomaly of the interior solution.
One can easily check that the trace (\ref{nrs}) is
proportional to $R_{\mu\nu\alpha\beta}R^{\mu\nu\alpha\beta}$. Indeed,
the trace anomaly in the Schwarzschild background is \cite{cf}
\be
\langle T^{\mu}{}_{\mu}\rangle=
\hbar\frac{3(a+b)}{\pi^2}\frac{(G_4 M)^2}{r^6}\,,
\ee
which comes entirely from the Riemann-squared term. According to our
discussion, the interior and exterior geometries considered in
\cite{bgm} cannot belong in the same theory, even if they were to be
both interpreted in the AdS/CFT context. In fact, using the
AdS$_5$/SYM/RS2 relation $L^2=(4/\pi)\hbar N^2 G_4$ in (\ref{nrs})
suggests that the exterior theory should have $a+b=-2N^2$. Obviously,
such matching is not physically sensible.\footnote{Away from the
horizon, the matching may be possible as a bubble at the interface
between the two phases. This might allow an interpretation of the
solutions in \cite{kt}.} Instead, one must look for a different
exterior, where the metric correctly encodes the quantum backreaction.
This naturally leads to a time-dependent evaporating black hole
(\ref{radmetric}).

Indeed, the matching to the far-field Vaidya metric (\ref{radmetric}),
is consistent with the anomaly check. The tracelessness of the
radiation stress-energy implies $R^\mu{}_\mu=0$, and so the anomaly
vanishes, with no contributions from the
$R_{\mu\nu\alpha\beta}R^{\mu\nu\alpha\beta}$ terms.  Although this
argument by itself does not fully guarantee that the bulk will be free
from singularities, it passes the anomaly check with only minimal
assumptions which are physically well-motivated.

Therefore, barring exotic possibilities, we see that the classical
bulk dynamics requires braneworld black holes to be time-dependent. We
have arrived at this conclusion by studying only the dynamics projected
on the $3+1$ braneworld, but we would also like to understand the
picture from the point of view of the full bulk AdS$_5$ spacetime. Then
the following questions arise naturally: {\it (i)} What is the bulk
dual of the Hawking radiation emitted by the black hole? {\it (ii)} Why
should a classical black hole on the brane {\it have to} emit anything?
{\it (iii)} Why should this emission, which is classical from the point
of view of the bulk, appear as {\it thermal} radiation in the dual 3+1
picture?

The answer to {\it (i)} is obvious: in the $3+1$ CFT+gravity
theory, Hawking radiation consists of CFT modes, whose dual in the
bulk are KK gravitons. The bulk emission consists of classical
gravitational waves. To answer {\it (ii)} we have to find a
natural mechanism that causes the black hole to classically emit
these waves into the bulk. Observe that the black hole is moving
along with the brane in AdS$_5$. The brane is a domain wall that
is accelerating away from the center of AdS. So the black hole
also accelerates, and as a consequence it must emit gravitational
waves. This means that the bulk dual of Hawking radiation is
gravitational bremsstrahlung. It would be interesting to
substantiate this qualitative idea with a more detailed analysis
of the relevant classical bulk physics. This will also shed light
on the important question {\it (iii)}, which for now is left open.
It is encouraging to note that we can at least reproduce the
estimate to leading order for the location of the peak of the
distribution, determining Hawking temperature, from purely
classical considerations in the bulk. Namely, the classical waves
which would be emitted into the bulk would have a characteristic
frequency determined by the inverse gravitational length of the
source, which for a $5D$ theory in the bulk is given by $\omega
\sim \sqrt{G_5 \rho}$, where $\rho$ is the energy density of the
region where the gravitational waves are emitted, $\rho \sim
M/{\rm Vol}_4$. Since the black hole is accelerating at a rate
$1/L$, we can estimate to leading order ${\rm Vol}_4 \sim r^3_H L
\sim G_4^3 M^3 L$, leading to $\rho \sim 1/(G_4^3 L M^2)$, and so
using (\ref{ans}) we find $\omega \sim 1/(G_4 M)$, i.e. precisely
the formula for Hawking temperature! However the complete
classical description of the thermal spectrum is yet to be
determined. Note that the bulk solution must be time asymmetric,
in contrast to the lower-dimensional solutions of \cite{ehm1},
where the black hole accelerates eternally and the net flux of
radiation vanishes.

Working on the bulk side, one should be able to reproduce the black
hole luminosity ${\cal L} \sim \hbar g_*/(G_4M)^2$ by solving classical
5D equations. Indeed, viewing the radiation loss as a classical effect
clarifies why this emission rate is so huge. It also explains why
the large release of energy into the bulk does not contradict the
statement that the black hole radiates mainly on the brane \cite{ehm3}:
this applied to Hawking radiation into the bulk, which was compared to
Hawking emission of non-CFT modes on the brane, in theories where there
may be additional degrees of freedom stuck to the brane. But from the
bulk point of view, the large bremsstrahlung emission we are
considering is {\it not} a quantum-mechanical process, and so is not
constrained by the analysis of \cite{ehm3}. It must not be confused
with Hawking radiation into the bulk, which is a much smaller effect.
From the dual CFT+gravity point of view, where the radiation is a
quantum phenomenon, the large black hole luminosity is simply a
consequence of the large number of CFT modes.

The bulk view would also allow one to follow the evolution of the
evaporating black hole beyond the threshold $r_H\sim L$, $M\sim
\sqrt{g_*}M_4$ at which the description in terms of a $3+1$ theory
of gravity+CFT breaks down, even down to $M_5 \ll M_4$, as we have
been arguing above. A black hole of size $r_H\ll L$ is
approximated near the horizon by a five-dimensional, {\it static}
Schwarzschild solution. Classical radiation into the bulk, and
therefore $3+1$ Hawking radiation into CFT modes, is suppressed
for such light black holes. An intuitive understanding of why this
happens may be gained from tunneling suppression \cite{dkkls}.
While large black holes are shaped like pancakes around the brane,
they extend to distances larger than the AdS radius $L$. Thus they
couple to all the CFT modes, including the lightest ones, with
$M_4$ couplings, without any suppressions. On the other hand,
while the small black holes are bulging away from the brane, they
are much smaller than the AdS radius, and from the perturbative
point of view, they live inside the RS2 `volcano'. Hence their
classical couplings to all bulk graviton modes are
tunneling-suppressed in the sense of \cite{dkkls}, and are
exponentially weaker than $M_4$. Thus the radiation rate must go
down significantly \footnote{The same effect occurs for a large
object on the brane, of mass $M\gg M_4$ but lower density than a
black hole, such as a star. Even if the star had accelerated by
being stuck to the brane, the bulk deformation it would cause
would have been confined to distances  less than $L$, so its
emission would have been tunneling-suppressed. The reason why a
black hole radiates in the bulk whereas a star does not is also
dual to the problem of the different choices of CFT vacua and
boundary conditions for the radiation. This deserves further
study.}. Hence the light black holes evaporate, although more
slowly, via bulk Hawking radiation. It is interesting to ask what
would be the bulk description of the final stages of black hole
evaporation. While on the CFT+gravity side, the classical area
theorems are violated by the quantum effects of Hawking radiation,
leading to the shrinking of the black hole horizon, on the bulk
side there are no quantum effect to leading order and the bulk
version of the area theorems still applies\footnote{We thank
D.~Marolf for very useful discussions on this issue.}. This would
imply that a black hole cannot disappear from the bulk. Hence a
consistent picture would be that the disappearance of a black hole
in the CFT+gravity theory corresponds to the classical sliding of
the black hole from the brane into the bulk. It would be very
interesting to verify this explicitly. Since this picture for the
evolution of an evaporating black hole is based on specific
properties of the UV extension provided by the bulk theory, there
is no reason why it should apply to situations that do not have an
AdS/CFT dual description.

\section{Conclusions}

We have proposed here a radical change of perspective on how to view
black holes in the context of AdS/CFT correspondence.
The previous work on black holes within the AdS/CFT framework has
been aimed at understanding a $D+1$-dimensional black hole sitting at
the center of AdS$_{D+1}$ in terms of the quantum states of a CFT at
the boundary. In this case, the black hole radiates via quantum effects
in the bulk, and one expects to learn about the quantum properties of a
black hole by studying its dual boundary description.

Instead, we put the black hole itself in the dual theory extended with
dynamical gravity. On the bulk side, this is realized by putting the
black hole on a brane in the cutoff AdS bulk, which localizes dynamical
gravity. Then we can study the quantum properties of a $D$-dimensional
black hole in terms of classical physics in the bulk. The quantum
Hawking radiation of CFT modes is described as the emission of
gravitational waves into the bulk, and the classical bulk point of view
may lead to a better understanding of quantum black hole evaporation.
Each of these two approaches prompts different classes of questions,
which can be naturally answered within these frameworks.

We have provided strong support for this new point of view with a
detailed analysis of the black hole solutions on a 2-brane in AdS$_4$
and their dual $2+1$ CFT+gravity description. Our analysis has also
revealed new features of the states of the $2+1$ CFT
coupled to $2+1$ gravity, and has shown explicitly that quantum effects
can censor singularities. We have found that the main properties of the
quantum censorship mechanism in $2+1$ dimensions are in fact quite
general, and should remain valid outside of the context of AdS/CFT. The
censorship is however amplified in the presence of many CFT modes, and
this appears to be the main requirement that makes the quantum censor
efficient.

In the context of the RS2 model in AdS$_5$, we have been able to
argue why an asymptotically flat, static, regular black hole
localized on the brane, could not be found. We emphasize
again that while we have been working in the context of AdS
braneworlds like RS2, which have proven to be a very useful tool
to study black holes, we expect that many of our results should
naturally extend to any CFT+gravity theory, even if a dual bulk
description along the lines of RS2 does not exist.

There remain a number of open issues. We have given a qualitative
argument for why a black hole on a brane should emit classical
gravitational waves, but it is still unclear why this emission,
which can be analyzed and understood in purely classical terms,
should project on the brane as a {\it thermal} flux of radiation.
The problem belongs to a class of connections between classical
effects in the bulk and thermal effects in the dual theory.
The conventional AdS/CFT approach tried to
understand how a state of the CFT encodes the classical causal
structure of the bulk black hole. The present problem is quite
different and could be an easier one, since we may have some hope
of analyzing the classical bulk physics involved in the radiation.

An aspect of our conjecture that we have only barely touched upon
is the choice of vacuum of the CFT. This is closely related to
understanding Hawking radiation as classical bulk bremsstrahlung.
It would be natural to expect that each consistent choice of vacuum
should correspond to a specific
bulk AdS solution, which differ from each other
by the boundary conditions for the bulk waves at
the bulk AdS horizons.
We have discussed a possible example in the case of BTZ
black holes. They admit both the $M=-1/8G_3$ and $M=0$ states as
consistent vacua, which we have conjectured to correspond to the two
branches of black holes localized on the brane. In $3+1$ dimensions
we also had alternative vacua, but we have only examined the physics
related to the Unruh vacuum, which models the late time behavior
of the collapse. The bulk dual of a black hole with backreaction
from the Hartle-Hawking state would be quite interesting as well:
The asymptotic thermal radiation is dual to a large black hole
inside the AdS$_5$ bulk. The motion of a brane in this spacetime
generates the radiation-dominated FRW evolution on the brane.
Hence the Hartle-Hawking state should be described in the dual
bulk theory as a black hole localized on a brane, which is itself
moving in the background of a large bulk black hole in the center
of AdS$_5$. The next-to-leading order corrections to the $2+1$
asymptotically flat black holes may lead to a similar picture. On
the other hand, the Boulware state should result in a null
singularity that is localized on the brane. It would be interesting to
check if there exists a relationship between these solutions and the
static linearized approximation in the RS2 model \cite{rs,gartan}.
We believe that these questions merit further consideration and hope to
return to them in the future.

\acknowledgments

We gratefully acknowledge useful discussions with D.~Amati,
R.~Balbinot, J.~Barb\'on, S.~Carlip, S.~Kachru, D.E.~Kaplan,
M.~Kleban, R.~Maartens, D.~Marolf, S.~Shenker, L.~Susskind and
T.~Tanaka. N.K. would like to thank the Yukawa Institute for
Theoretical Physics at Kyoto University for hospitality while this
work was being initiated.  The work of R.E. has been partially
supported by CICYT grant AEN-99-0315. The work of N.K. has been
supported in part by NSF Grant PHY-9870115.

\end{document}